\documentclass[manuscript,nonacm]{acmart}

\setcopyright{none}

\usepackage{url}
\usepackage{hyperref}

\AtBeginDocument{%
  }

\begin{document}
\thispagestyle{empty}
\vspace*{-2em}
\small
\begin{flushright}
    \textbf{CHIWORK 2025 Workshop on Generative AI Disclosure, Ownership, and Accountability in Co-Creative Domains} \\
    \rule{\linewidth}{0.1pt}
\end{flushright}
\vspace{1em}
\title{Penalizing Transparency? How AI Disclosure and Author Demographics Shape Human and AI Judgments About Writing}

\author{Inyoung Cheong}
\affiliation{%
  \institution{Princeton University}
  \city{Princeton}
  \state{NJ}
  \country{USA}
}

\author{Alicia Guo}
\affiliation{%
  \institution{University of Washington}
  \city{Seattle}
  \state{WA}
  \country{USA}}

\author{Mina Lee}
\affiliation{%
  \institution{University of Chicago}
  \city{}
  \state{WA}
  \country{USA}}

\author{Zhehui Liao}
\affiliation{%
  \institution{University of Washington}
  \city{Seattle}
  \state{WA}
  \country{USA}
}

\author{Kowe Kadoma}
\affiliation{%
  \institution{Cornell University}
  \city{Ithaka}
  \state{NY}
  \country{USA}}


\author{Dongyoung Go}
\affiliation{%
 \institution{Naver}
 \city{Bellevue}
 \state{WA}
 \country{USA}}

\author{Joseph Chee Chang}
\affiliation{%
  \institution{Allen Institute for AI}
  \city{Seattle}
  \state{WA}
  \country{USA}}

\author{Peter Henderson}
\affiliation{%
  \institution{Princeton University}
  \city{Princeton}
  \state{NJ}
  \country{USA}}

\author{Mor Naaman}
\affiliation{%
  \institution{Cornell Tech}
  \city{New York}
  \state{NY}
  \country{USA}}

\author{Amy X. Zhang}
\affiliation{%
  \institution{University of Washington}
  \city{Seattle}
  \state{WA}
  \country{USA}}

\renewcommand{\shortauthors}{Cheong et al.}

\begin{abstract}
  As AI integrates in various types of human writing, calls for transparency around AI assistance are growing. However, if transparency operates on uneven ground and certain identity groups bear a heavier cost for being honest, then the burden of openness becomes asymmetrical. This study investigates how AI disclosure statement affects perceptions of writing quality, and whether these effects vary by the author's race and gender. Through a large-scale controlled experiment, both human raters (n = 1,970) and LLM raters (n = 2,520) evaluated a single human-written news article while disclosure statements and author demographics were systematically varied. This approach reflects how both human and algorithmic decisions now influence access to opportunities (e.g., hiring, promotion) and social recognition (e.g., content recommendation algorithms). We find that both human and LLM raters consistently penalize disclosed AI use. However, only LLM raters exhibit demographic interaction effects: they favor articles attributed to women or Black authors when no disclosure is present. But these advantages disappear when AI assistance is revealed. These findings illuminate the complex relationships between AI disclosure and author identity, highlighting disparities between machine and human evaluation patterns.
\end{abstract}



\maketitle

\section{Transparency Demands and Concerns about Stigmatization}

AI systems are widely integrated into writing workflow as information sources~\cite{shah2022situating, buschek2024collage}, readily available proofreaders~\cite{liao2024llmsresearchtoolslarge}, or ``thought partners''~\cite{Collins_2024, kim2024diarymate, wan2024felt, benharrak2024writer}. This transformation shifts our expectations surrounding authorship, originality, and intellectual labor~\cite{buschek2024collage, lee2024design}. Readers, editors, and reviewers increasingly call for clear disclosure of AI involvement~\cite{boyd2023acl, cho2023papercard}: Did the machine write this? How much of it? Can I still trust it? Beneath these questions lies a desire to calibrate judgments to discern the boundary between human insight and synthetic fluency. 
\begin{figure*}[ht]
    \centering
    \includegraphics[width=\linewidth]{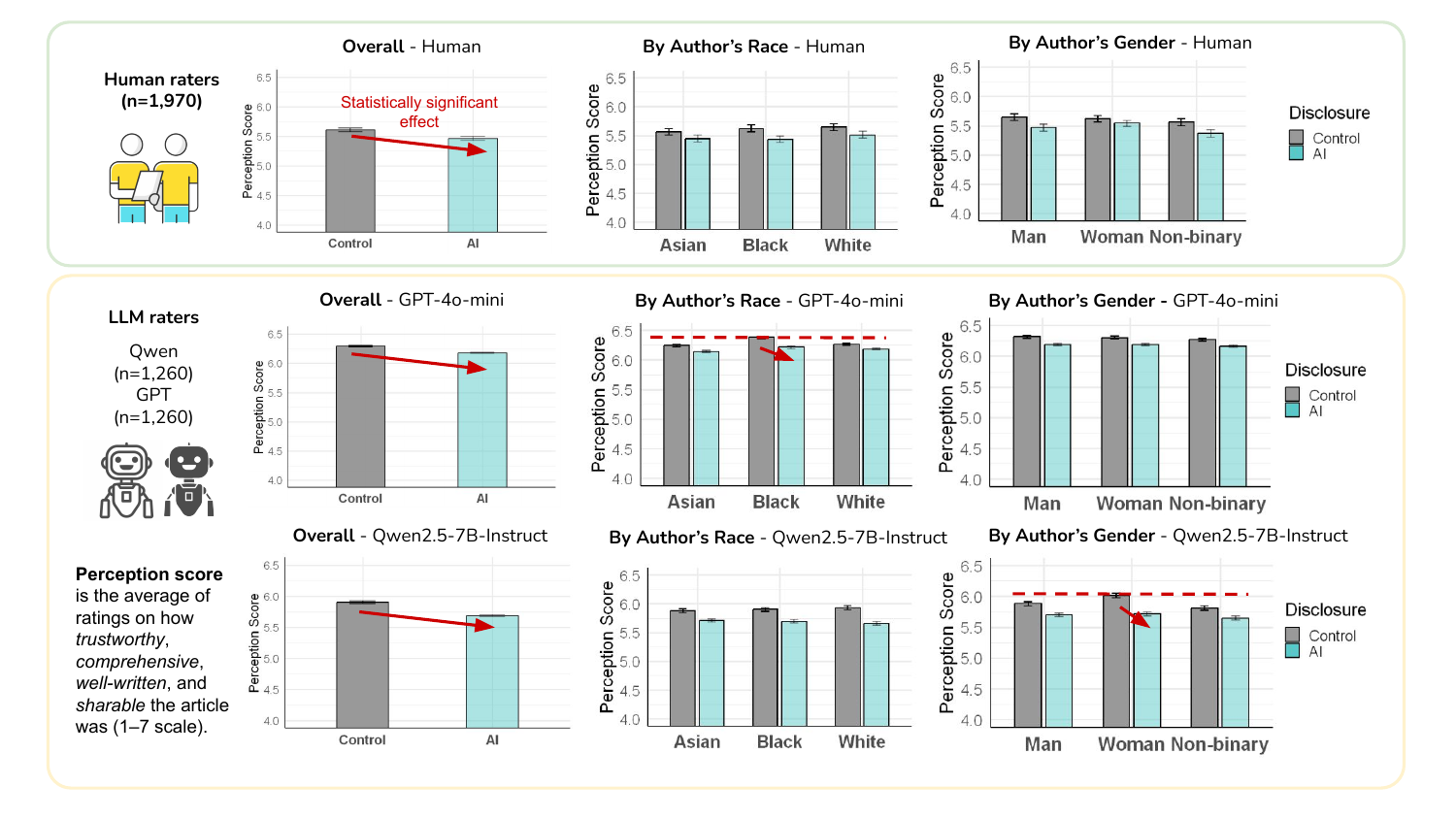}
    \caption{Perception scores across 9 evaluation conditions, comparing human raters (top row) with large language models GPT-4o-mini and Qwen2.5-7B-Instruct (middle and bottom rows). Each panel shows mean perception score of raters with and without AI disclosure, across author's demographic groups. Human participants consistently penalized disclosed articles, but showed no demographic interaction effects. In contrast, GPT-4o-mini exhibited favoritism toward Black authors and Qwen favored Woman authors—but only in the absence of AI disclosure. When AI assitance was disclosed, these demographic preferences were largely eliminated.}
    \label{fig:chart}
\end{figure*}

Transparency, while noble in theory, is complicated in practice. Writers may hesitate to disclose their use of AI tools because they fear how their work will be perceived. Indeed, Baek et al.~\cite{baek2024effect} found that labeling identical content as AI-assisted can reduce perceptions of credibility, creativity, and shareability. We wondered: Does this ``AI discount'' fall equally on all authors? Audiences do not assess writing in a vacuum; perceptions are shaped by the perceived identity of the author. For instance, Klaas \& Boukes~\cite{klaas2022woman} found that male journalists were considered more credible than their female counterparts, particularly when reporting on technology. Similarly, Eom et al.\cite{eom2025race} showed that Black male scientists were perceived as warmer, while White female scientists were rated as more competent. 
If reader perceptions about AI disclosure are filtered through gendered, racialized, or other socio-epistemic cues, then the burden of openness becomes asymmetrical. Researchers conceptualize this negative societal judgments associated with someone's perceived or actual use of AI systems as ``perceptual harms.''~\cite{kadoma2024generative} With these harms, the groups historically under-resourced and under-represented face the greatest risk of stigmatization for using AI.

Moreover, humans are no longer the sole evaluators of writing. Institutions are turning to AI to process large volumes of writing quickly, reduce costs, and sometimes, pursue the appearance of neutral judgment. In Texas, public school essays on standardized tests are now graded by AI systems trained to mirror human scorers, sparking concern among teachers and parents about fairness and accountability~\cite{Peters_2024}. An estimated 99\% of Fortune 500 companies use some form of automated screening in hiring~\cite{Hanson_2023}. AI-based performance review systems have been promoted to evaluate employee data in real-time and offer feedback based on actual metrics rather than human bias or memory~\cite{People_Rice_2024}. In the publishing world, traditional publishers have started paying attention to metrics and even hiring data scientists to spot underrated gems that human editors might overlook~\cite{Althoff}. Inkitt, a self-publishing platform, uses machine learning to evaluate user-submitted novels and identify breakout titles, guiding editorial investment and promotion~\cite{Lunden_2024}. These systems have become arbiters in high-stakes decisions, shaping whose writing is seen, trusted, or excluded.

This motivates us to investigate a question: \textbf{When a writer discloses AI use, do readers judge the writing differently based on who the writer is?} To answer this question, we designed a large-scale experiment combining both human and LLM raters. Human participants (n = 1,970) evaluated the same news article, randomly assigned to one of eighteen conditions varying in author race, gender, and the presence of an AI disclosure statement. To compare how machines might reproduce or deviate from human judgment, we also evaluated the same content using two state-of-the-art vision-language models: GPT-4o-mini and Qwen2.5-7B-Instruct. This combined approach allows us to assess not only how AI disclosure affects perceptions of writing, but whether those effects depend on author's identity, according to both humans and machines.

\section{Human Participants Penalized AI-disclosed Articles across All Demographic Groups}

We conducted a pre-registered, large-scale survey with 1,970 participants.\footnote{The pre-registration is available at \url{https://osf.io/2wbm9}.} The study employed a 2×3×3 between-subjects factorial design with three independent variables: (1) \textbf{AI disclosure} (presence or absence of a disclosure statement), (2) \textbf{author race} (Asian, Black, White), and (3) \textbf{author gender} (man, woman, non-binary). Participants were randomly assigned to one of eighteen experimental conditions and evaluated an identical news article, with only the author biography and disclosure language varying across conditions. The two types of disclosure statements used in the study are shown in Table~\ref{tab:disclosure}. Additional details on the rationale behind this research design are provided in Appendix~\ref{app:design}.

\begin{table}[h]
\centering
\caption{Disclosure statements shown to participants.}
\begin{tabular}{ll}
\toprule
\textbf{Condition} & \textbf{Disclosure Statement} \\
\midrule
Control & \textit{Statistical information updated as of Oct. 11, 2024.} \\
AI      & \textit{This article was created with assistance from Artificial Intelligence (AI) tools.} \\
        & \textit{Statistical information updated as of Oct. 11, 2024.} \\
\bottomrule
\end{tabular}

\label{tab:disclosure}
\end{table}

The study has four phases. In Phase 1, participants were exposed to identical, human-authored news article across all conditions, with variations in the author biography, photo, and disclosure statement (\textit{See} Appendix \ref{app:survey-interface}). The control condition includes a basic statistical update disclosure, while the treatment condition explicitly states the use of AI tools in content creation. During Phase 2, participants engaged in a deception task where they identify evidence and recommendations from each paragraph. Phase 3 consisted of dependent measures where participants rate four key aspects using 7-point Likert scales (1 = Strongly Disagree, 7 = Strongly Agree): information trustworthiness, comprehensiveness, writing quality, and likelihood of sharing. In Phase 4, participants filled an exit questionnaire capturing recall of the disclosure statement and author demographics, personal and professional AI usage patterns. 

To examine whether author identity moderated the AI disclosure effect, we fit a series of linear models with interaction terms. In the baseline model, articles with an AI disclosure received significantly lower ratings compared to those without disclosure ($p = 0.021$), \textbf{confirming the presence of a disclosure penalty}. However, while the AI disclosure penalty persisted across demographic subgroups, the interaction between disclosure and author identity was not significant. This finding suggest that, while the disclosure of AI involvement leads to a modest reduction in article evaluations, this effect operates \textbf{relatively independently of the author’s perceived race or gender}.  

\section{LLM Raters Exhibit Demographic Preferences But Suppress Them When AI Disclosure was Present} 

We extend this inquiry by evaluating the same news article across the same eighteen conditions using two state-of-the-art vision-language models: GPT-4o-mini and Qwen2.5-7B-Instruct. Each model generated 1,260 evaluations (one per condition and prompt combination), for a total of 2,520 model ratings. Both models accept text and image inputs, allowing them to process both article content and author photos simultaneously. The results suggest that these AI evaluators exhibit complex patterns of bias when evaluating articles with AI disclosure. 

First of all, both models presented statistically significant AI disclosure penalty (-0.112 points, $p = 0.002$ for GPT-4o-mini and -0.133 points, $p = 0.026$ for Qwen2.5-7B-Instruct). However, unlike humans, the AI models demonstrated distinct demographic preferences in the control condition. GPT-4o-mini showed significant favoritism toward Black authors (+0.137 points higher than Asian authors, $p < 0.001$), while Qwen demonstrated a preference for woman authors (+0.133 points higher than man authors, $p = 0.004$). Interestingly, both models \textbf{penalized their preferred demographic groups more heavily} when AI assistance was disclosed. Post-hoc analysis using Tukey's adjustment confirmed that while demographic differences were significant in the control condition, they largely disappeared in the AI-disclosed condition. For example, in GPT-4o-mini's evaluations, Black authors scored significantly higher than Asian authors in the control condition ($p < 0.0001$), but this difference was barely significant in the AI disclosure condition ($p = 0.0435$). Similarly, for Qwen, woman authors scored significantly higher than man and non-binary authors in the control condition ($p < 0.0001$ and $p = 0.0113$ respectively), but these differences became non-significant when AI assistance was disclosed.

\section{Disclosure, Author's Identity, and the Fragility of LLM Alignment} 

Our two-pronged investigation shows that humans and LLM raters alike apply a modest but consistent penalty to articles when AI assistance is disclosed. This finding shows that AI disclosure carries epistemic stigmatization, although the magnitude of ``AI disclosure discount'' is small—less than 0.15 on a 7-point scale across all experiments, indicating a perceptible yet not overwhelming penalty. This finding aligns with prior work showing that AI disclosure has a statistically significant but relatively modest impact on perception~\cite{Lim_Schmälzle_2024}.  

However, a key divergence emerged: interaction effects between author identity and disclosure status were present only in LLM evaluations. Among human raters, disclosure penalties were relatively uniform across authors of different races and genders. In contrast, both LLMs exhibited identity-disclosure interaction effects. GPT-4o-mini showed a pronounced preference for Black authors in the control condition, which diminished when AI involvement was disclosed. Qwen2.5-7B-Instruct similarly favored woman authors in the absence of disclosure, a gender bias that disappeared when AI assistance was acknowledged.

Notably, both favored groups—Black authors and women—are historically marginalized. These preferences may reflect an alignment-driven over-correction, in which models trained with human feedback disproportionately reward underrepresented identities~\cite{kumari2024dynamic}. However, when AI disclosure is present, these fairness-oriented preferences vanish. We term this dynamic a form of ``vanishing alignment,'' where the social and ethical calibration of model behavior becomes fragile under changing contextual cues. A similar phenomenon has been observed by Hofmann et al.~\cite{hofmann2024ai}, who found that LLMs trained with human preference alignment projected overtly positive stereotypes toward African American English speakers while maintaining covertly negative biases that surfaced in indirect prompts. Our findings suggest that similarly, fairness-aligned behaviors in LLMs may be conditional and unstable.

Several factors may help explain why these effects appear in LLMs but not in human ratings. While human raters bring diverse lived experiences and norms to the task, LLM raters might produce more stable, patterned outputs across multiple runs, making their biases more legible. Moreover, the genre of the writing (news articles) may also play a role~\cite{wan2024coco}. AI disclosure may trigger genre expectations around objectivity, amplifying penalties for AI involvement and reducing the salience of the author's race or gender. Our research offers groundwork for understanding how AI disclosure shapes perceptions of writing and author's identity. Future work should explore the epistemic consequences of AI disclosure across genres, audiences, and forms of AI involvement and critically, attend to how perceptual harms may accumulate or compound in various evaluative settings.

\bibliographystyle{ACM-Reference-Format}
\bibliography{sample-base}

\appendix

\section{Survey Interface}

\label{app:survey-interface}
Our survey interface is available at \href{https://social.cs.washington.edu/news-annotation/health/}{https://social.cs.washington.edu/news-annotation/health/}. Below we present the side-by-side screen participants saw in our web-based experiment: (a) the article display with author profile and AI disclosure, and (b) the questionnaire page with 7-point Likert items.

\begin{figure}[h!]
    \centering
    \includegraphics[width=\textwidth]{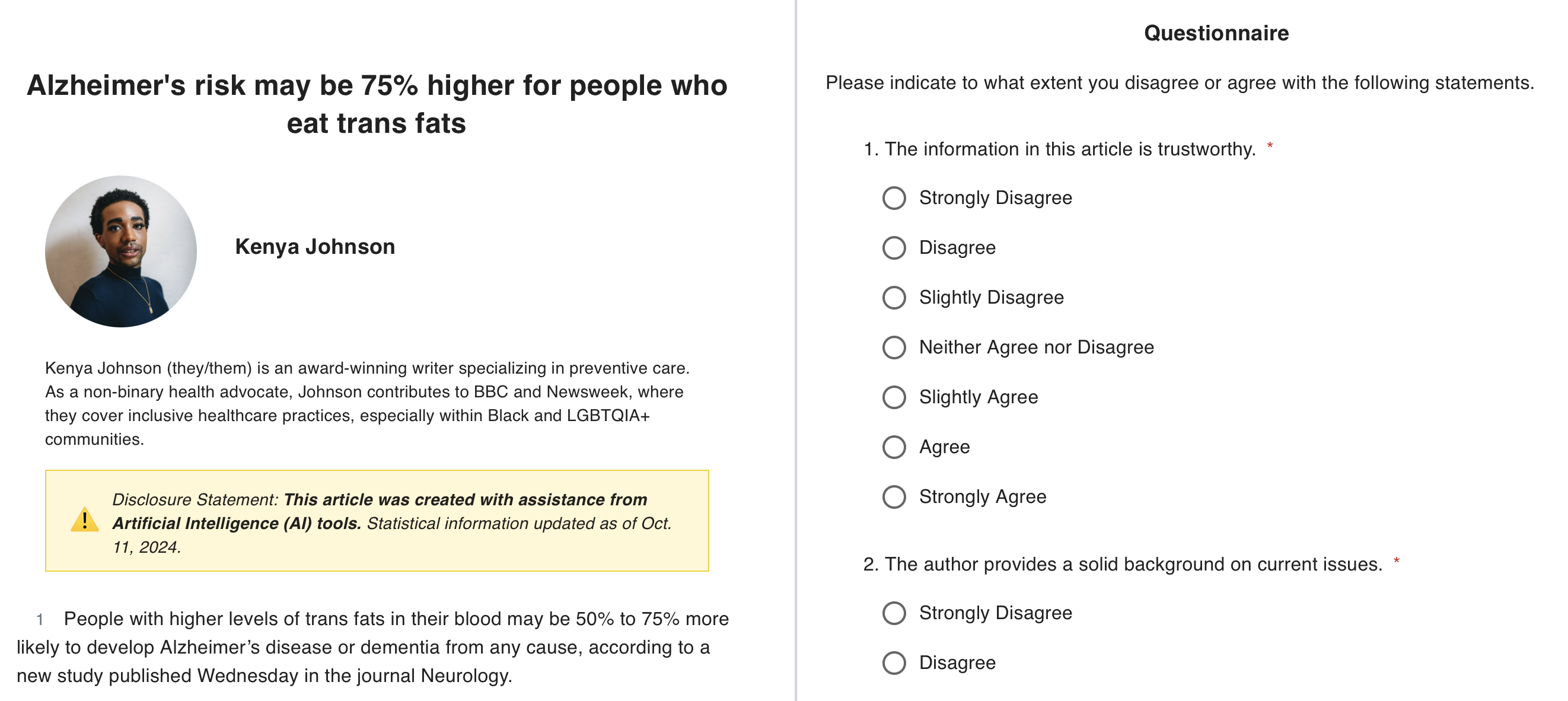}
    \caption{(a) Participants read a health‐news article with author photo, name, pronouns, bio, and the AI‐disclosure banner. (b) While reading the article, they rated trustworthiness, comprehensiveness, quality, and share likelihood on a 7‐point Likert scale.}
    \label{fig:survey-questionnaire}
  \label{fig:survey-interface}
\end{figure}

\section{Human Evaluation Study Design}
\label{app:design}

The biggest challenge in bias assessment surveys is that participants who realize their biases are being measured might deliberately assign higher scores to minority authors, undermining result reliability. To address this, we did multiple iterations on a survey environment to elicit candid feedback without revealing our purpose. Our initial within-subjects design, where each person evaluated multiple articles varying by author demographics, quickly alerted participants to the bias assessment in a pilot study. We subsequently switched to a between-subjects design, presenting each participant with only one article from one of eighteen demographic subgroups. This eliminated direct comparisons between different authors and articles and required recruiting over 2,000 participants to ensure adequate sample sizes across all conditions.

Presenting author demographics effectively posed another challenge. Our original race categories included Asian, Black, Hispanic, and White authors. However, even after adding more detailed demographic descriptions such as a location and an expertise area, misclassification rates for Hispanic authors remained high. Therefore, we narrowed our racial categories to (East) Asian, Black, and White, with more salient visual cues. In addition, although our initial design did not include detailed author biographies, following standard presentation of news articles. However, pilot testing showed that photos and names were not good enough to remain impressions to participants. We therefore enhanced demographic signals by adding pronouns and identity-related expertise (e.g., inclusive healthcare practices within Black and LGBTQIA+ communities). This created a tension between making demographic features salient enough to be noticed but diverting from industry standards risked triggering over-corrective behavior from participants who might detect the study purpose.  

The final design succeeded in concealing the real purpose of the study. Only 1.1\% of participants correctly guessed we were examining demographic bias. Our manipulation checks confirmed this effectiveness, with 73.4\% correctly recalling author gender, 78.4\% recalling author race, and 78.8\% noting the presence or absence of AI disclosure. 52.7\% correctly recalled all three elements. Overall, this study lays the groundwork for a systematic understanding of AI disclosure effects.

\section{Data Availability}
\label{app:data}
The survey data, LLM evaluation outputs, and analysis code are available at \url{https://osf.io/shx4v/?view_only=d4f876ca6ae44c0cb840d8e2230d62e5}. 

\end{document}